# Optimizing Connectivity and Scheduling of Near/Far Field Users in Massive MIMO-NOMA System

Ziad Qais Al-Abbasi

**Abstract:** It is envisioned that the next generations of wireless communication environment will be characterized with dense traffic demand due to the prediction that there will be large numbers of active users. Hence, it is important to find a solution to deal with such dense numbers of users. This paper investigates optimizing the connectivity and users scheduling to improve the performance of near and far field users in a downlink, multiuser, massive MIMO-NOMA system. For the considered system model, combining NOMA side by side with massive MIMO offers a great opportunity to exploit the available radio resources and boost the overall system efficiency. The paper proposes separate clustering of near field users and far field users. It also proposes using a beamforming scheme to separately serve the users within each cluster. However, NOMA is proposed to be applied among all users to boost resource sharing. In particular, a cognitive-NOMA beamforming scheme and NOMA themed beamforming are proposed to serve the users within each cluster, and they are compared against random beamforming from literature. Simulation results show that both of the proposed beamforming schemes proved their superiority as compared to random beamforming. Several scheduling techniques were also considered in this paper to examine possible solutions for boosting the system performance considered, namely, priority, joint, dynamic, and fairness-based scheduling techniques for both near field and far field users. The paper also proposes a suboptimal, fairness aiming and gradual allocation approach for allocating the transmission power among the users. The results show that user-clustering offers better connectivity and scheduling performance than the case where no clustering is applied.

Middle Technical University (MTU), Baquba Technical College (BTC), Baghdad 10074, Iraq. ORCID: 0000-0001-7062-3383 (Email: ziad.al-abbasi@mtu.edu.iq).

**1 Introduction**

Due to the expected dense traffic demand in future wireless networks, segregating users into smaller clusters leads to better management of resources and better resources utilization, especially with the existence of NOMA as it optimizes EE, sum rate, and overall network performance [1]. One of the key technologies to 5G physical layers is the massive MIMO technique. Massive MIMO concept is to adopt a large number of antennae at the base station to achieve spatial multiplexing among a large number of users over the same time/frequency resource, hence, it offers large antenna array gain, and it means the communications will occur within the near field range [2], [3]. In addition, operating at high frequency along with the deployment of large number of antennas also lead to the communication occurrence to be in the near field region [4]. The proposed work in this paper encompasses these concepts, hence, the considered system model will have a near field and far field regions set up. It is anticipated that 6G networks would be designed to support an unprecedented number of users and handle diverse range of communication requirements. Therefore, it is necessary to establish effective user management strategies for such purposes. In massive MIMO-NOMA systems, where multiple users share the same resources, due to varying distances from the base station, near and far-field mobile users experience significantly different channel conditions. If not tackled effectively, this disparity can lead to uneven resource allocation and cause detrimental impact upon system efficiency in general. Managing users through user clustering, which groups users with identical channel characteristics, offers a powerful approach to address this issue. Users' clustering facilitates optimizing connectivity and scheduling, optimizing resource allocation, minimizing inter-cluster interference, and enhancing the spectral efficiency of the system. This targeted clustering approach not only simplifies the implementation of NOMA and massive MIMO directive transmission, but also improves coverage, users' scheduling, and connectivity for both near and far-field users. All of the above arguments make users' clustering a vital strategy to be exploited for advancing massive MIMO-NOMA systems in future 6G networks. The contribution provided in this paper could be summarized as follows. For the near field users, this paper proposes adopting priority scheduling versus dynamic scheduling. On the other hand, for far field scheduling, this paper proposes fairness-based scheduling versus joint scheduling. The performance of all those scheduling schemes were examined with the proposed beamforming schemes, namely, NOMA and cognitive-NOMA beamforming schemes. To cope up with the expected challenges that are predicted to arise with 6G, this paper also proposes clustering the users to facilitate serving them with the relevant level of quality of service.

1.1 Related Works
A considerable literature works has proven the efficient potential of massive MIMO and NOMA in offering high capacity, energy efficiency, and sum rate under predicted set up that is close to the 6G of wireless systems. Several works have proposed applying user pairing [5], [6], [7], [8] and user clustering [9], [10], [11], [12], [13], [14], [15], to facilitate serving many users within a future wireless communication system. The differentiations between the two approaches could be summarized as illustrated in Table I.

**Table I:** The differences between the concepts of clustering and pairing.

| Clustering | Pairing |
| --- | --- |
| Each cluster occupies several users. | Occupies only two users |
| Gathering strategy could depend upon similar channel powers, nearby CSI, spatially approximate users, and so forth. | Depends upon contradicted features such as high channel power is paired along with low channel powered user. |
| Inter cluster Interference management as well as intra cluster interference management | Interference management is dealt with through SIC |
| The users within each cluster could benefit from joint beamforming. | Beamforming applied specifically to the users within each pair. |
| Suitable for scenarios where massive MIMO is applied, and spatial domain is exploited. | Suits scenarios such as NOMA assisted system and users with channel gaps. |

However, in this paper, an investigation on the effect of users' clustering is applied, at first, to check if it's convenient to be applied upon a near/far field users communicating through the downlink of massive MIMO-NOMA system. The near/far field communications have attracted considerable attention where several aspects have been examined so far, however there is still more to come in this field as it is expected to play a major role in the next generations of wireless communications systems. For instance, vital aspects of near/far field communications are still yet to be examined in detail with proper specification of the expected challenges and the possible solutions for 6G scenarios. In particular, the users' scheduling is predicted to be challenging under the pressure of dense active users and the demand for variety of services and traffic. In addition, establishing reliable connectivity in near/far field communications is also not well examined in literature. This paper focuses on discussing such problems and proposes solutions for them. Another research spot to be examined is the beam focusing especially on scenarios that would exploit massive MIMO. For such purposes, this paper also proposes cognitive beamforming as a solution for systems that adopt MIMO-NOMA to serve communicating users in next generations of wireless networks. Other aspects were determined by a few works of literature. The boundaries of the near field and that of the far field were determined by the authors of [2], [16]. They also highlighted the main challenges and features related to the near field communications, such as channel modeling and estimation, beamforming design and hardware manufacturing. The massive MIMO behavior in the near field region is studied in [9]. In addition, the near field of holographic MIMO is also studies in [17]. A dynamic architecture of MIMO for near field communication is designed in [18]. The authors exploited the near field channel to propose a hybrid beamforming design for future applications. The channel modeling of a near field holographic MIMO is presented in [19] where the authors explored an overview of the possible evaluation metrics and the main challenges of near field channel models. A review about the near field communication characteristics related to reconfigurable intelligent surfaces (RIS) based 6G networks is presented in [20]. Where the author highlighted the necessity to establish a practical near field propagation environment, RIS networks, building a near field paradigm for 6G networks, and highlighting the main challenges and obstacles that would face future networks in the near field region. The authors of [21] investigated the near field communications characteristics for a holographic MIMO system. They stated that the channel correlations at the far field region are higher than those at the near field region which results in lowering the achievable capacity. The capacity issue will be taken into consideration in this paper to optimize the capacity and the experience of the far-field users. The work in this paper is also an additional effort to respond and complete what has been mentioned in the literature works. The coexistence of millimeter-wave (mmWave) communications and non-orthogonal multiple access (NOMA) technologies is promising for future communications. Hence, this combination has drawn considerable attention to literature along with supporting schemes such as random beamforming that could be adopted in massive MIMO systems similar to the one considered in this paper. Random beamforming removes the necessity that the BS acknowledge the channel state information (CSI) of the users as it exploits stochastic geometry to assist in characterizing the transmission. For instance, in [22], the authors considered MIMO-NOMA in mmWave communications systems. In which an optimal power allocation scheme is proposed along with a random near-random far (RNRF) technique. The results refer to an important reduction in the outage probability gained by adopting the proposed random beamforming technique. Therefore, in this paper, the random beamforming will be adopted for comparison and to check its performance for massive MIMO rather than MIMO. In addition. The authors of [23] also highlighted the effectiveness of random beamforming by considering it in mmWave-NOMA systems with sparse-antenna array. They stated that random beamforming is vital for improving the connectivity and achieve low latency in mmWave-NOMA system. Random beamforming was also proposed in [24] with mmWave-NOMA system, and in [25] it was proposed to be adopted with MIMO-NOMA system. This motivates us to include random beamforming in comparison against the proposed approaches in this paper to check the superiority of those schemes and their potential to serve the future communication systems.

## 2. Motivation and Main Contributions

Future 6G wireless communication is predicted to support communications and services not only on ground, but also in space, sea, and air as well. In addition, it is supposed to offer variety of functionalities beside end-to-end communications, such as, sensing, positioning (e.g., GPS), computing,…etc. such features require the wireless systems to be equipped with advanced technologies and capabilities to ensure that all of those features are performed by the system while maintaining high throughput, occupying massive number of connections, spectral efficiency, high sum rate, energy efficiency, low latency and efficient resource utilizations. Motivated by the above statements in addition to the predicted demands accompanying the next generations of wireless communication systems; this paper proposes applying user-clustering to near/far users in massive MIMO-NOMA system. The goal behind clustering the users is to facilitate dealing with the expected traffic demand, various services quality at the receiving end, efficient radio resources usage, boosting the connectivity level through proposing suboptimal and optimal power allocation schemes. In addition, increasing energy efficiency through optimized power allocation and reduced power consumption. This paper also proposes a suboptimal fairness-gradual (FG) power allocation scheme that allocates the transmission power among the users while benefiting from clustering. The proposed scheme is compared against optimal power allocation strategy in terms of energy efficiency and sum rate.

## 3. The Investigated System Model

There are a number of practical implications that are considered while setting up the system model to be adopted in this paper while demonstrating the applicability and relevance of the proposed methods in this paper so that those methods are adopted for future 6G systems. Those implications are summarized as follows:

• Deployment Feasibility: the proposed connectivity and scheduling schemes are designed while keeping in mind the constraints of 6G hardware. This includes the antenna size of massive MIMO array and the computational traffic demands. These proposed approaches can be implemented with real-time processing units that are being gradually adopted in B5G and 6G systems.
• Interference Management: The proposed cognitive beamforming scheme effectively removes the interference between near and far-field users. This is vital for environments such dense urban areas with high device densities, and variety of service demands as expected in 6G systems. The considered system model in this paper leverages NOMA and massive MIMO together to establish robust outcome in terms of connectivity and SINR under these challenging conditions.
• Network Scalability: The proposed schemes are scalable in terms of various numbers of users and massive MIMO antennas. They could handle different network sizes, from small cells to large macro deployments. This is crucial for 6G as those networks are expected to support variety of applications, from massive IoT connectivity, scheduling large numbers of users, and up to high-data-rate services.
• User Mobility: In highly dense and dynamic environments that are characterized with mobile users, the proposed techniques for scheduling and connectivity can adapt to the user's movement, while guaranteeing consistent service for near and far-field users. The scheduling algorithms are flexible and can occupy real-time mobility challenges that are typical in 6G deployments.

The chosen system model in this paper is set up in a way to tackle the above discussed practical implications. This paper studies a system model that consists of a downlink of near/far field users in a massive MIMO-NOMA system. The base station is considered to be equipped with massive MIMO antenna technology to serve several single antenna users. The propagation environment is assumed to be characterized with Rayleigh flat fading channel is considered to affect all the users as well as other propagation environment effects, such as the distance between the users and the BS, the mobility nature of the user which is considered as a random variable to take several cases including moving and static users. The level of traffic demand is also taken into account while counting for the channel effects to guarantee that the results obtained are close to practical scenarios. To classify the user as being in a near field or far field, their distance from the serving base station is measured and compared against the Rayleigh distance $D^{Rayleigh}$, which in turn expressed as given in equation (1) [26], [27].

$$D^{Rayleigh} = \frac{2((N-1)d)^2}{\lambda} \quad (1)$$

Where $N$ represents the number of massive MIMO antennas at the base station, $d$ denotes the distance between each user and the base station, and $\lambda$ refers to the wavelength. It is worth noting that the expression in equation (1) is applied to all the $U$ users to classify the users into near field and far field users. The reference Rayleigh distance is set as $D_0^{Rayleigh} = \frac{2((N-1)d_0)^2}{\lambda}$, where it is calculated at the reference distance value $d_0 = 20m$. Users whom their Rayleigh distance found to be higher than $D_0^{Rayleigh}$ will be classified as far field users. On the other hand, users with Rayleigh distance that is lower than $D_0^{Rayleigh}$ will be regarded as near field users. Where the total number of near field users will be termed as $U^{near}$ and that of the far users will be $U^{far}$, keeping in mid that $U^{near} + U^{far} = U$.

## 3.1 Applying NOMA Among Near-Field Users

NOMA is known to improve the EE, SE, and the capacity of single antenna and multiple antenna-based systems [28], [29], [30]. Therefore, it is possible to benefit from it by clustering the near-field users as they are characterized by relatively homogeneous propagation environment due to the small size of their premises. Where the users with high channel gains could be clustered together to share the same radio resources simultaneously and different power levels allocated to those users. In such case, the efficiency of resource utilization will improve and there will be extra capacity to accommodate additional users in the system. In a downlink massive MIMO-NOMA system, the received signal by the $i$-th user under a small-scale Rayleigh fading effect, is expressed as given in equation (2).

$$y_{near}^i = h_{near}^i \left( \sqrt[2]{p_{near}^i} s_{near}^i + \sum_{m \neq i}^{U} \sqrt[2]{p_{near}^m} s_{near}^m \right) + n_{near}^i \qquad (2)$$

where $h^i$, $p^i$, $s^i$ and $n^i$ denote the channel power between the $i$-th user and the BS, the signal intended for the $i$-th user, transmission power allocated to the $i$-th user, and the additive white gaussian noise (AWGN) affecting the $i$-th user, respectively. The near field channel model is represented by the spherical coordinates system in accordance with spherical propagation model, and its vector is given by equation (3) [31].

$$h_{near}^i = \eta_{near}^i \left[ exp^{-j\frac{2\pi}{\lambda}d_{near}^i} \dots exp^{-j\frac{2\pi}{\lambda}d_{near}^M} \right]^T \qquad (3)$$

Where $\eta_{near}^i = \frac{c}{4\pi f_c d_{near}^i}$, the terms $c$ and $f_c$ denote the speed of light in free space and the operating carrier frequency, respectively.

At the receiver, NOMA users apply successive interference cancellation (SIC) mechanism to decode their data signal. According to SIC, the $i$-th user will perform SIC to remove the users with lower channel powers before decoding its intended signal, on the other hand, it will treat the users with higher channel powers as noise. Hence, the received signal to interference plus noise ratio of the $i$-th user $\Gamma_{near}^i$ could be expressed as in equation (4).

$\Gamma_{near}^i$

Then the achievable sum rate $\zeta$ is simply determined by equation (5).

$$\zeta_{near} = \sum_{i=1}^{U} \log_2(1 + \Gamma_{near}^i) \qquad (\frac{bit/sec}{Hz}) \qquad (5)$$

## 3.2 Applying NOMA among far-field users

In the case of far-field users, NOMA could be applied at the areas between the cells to enhance the level of the coverage provided as well as enhance the spectrum sharing among those users. The received signal by each far user is expressed in equation (6).

$$y_{far}^i = \frac{h_{far}^i}{d^s} \left( \sqrt[2]{p_{far}^i} s_{far}^i + \sum_{m \neq i}^{U} \sqrt[2]{p_{far}^m} s_{far}^m \right) + n_{far}^i \qquad (6)$$

Where $d^s$ represent the distance between the $i$-th user and the base station and the upper script $(\ )^s$ denotes the path loss exponent. The received signal to interference plus noise ratio of the $i$-th user $\Gamma_{far}^i$ could be expressed as depicted in equation (7).

$$\Gamma_{far}^i = \frac{p_{far}^i \left( |(h_{far}^i)^H w_{far}^i|^2 / d^s \right)}{\sum_{m \neq i}^{U} p_{far}^m \left( |(h_{far}^i)^H w_{far}^m|^2 / d^s \right) + \sigma^2} \quad (7)$$

The expression in equation (7) encompasses that, due to the application of NOMA, the far field users will experience interference from all other users because their signals are decoded as a first step. Then comes the stage of decoding the near field users signal as could be seen in equation (4). Then the achievable sum rate $\zeta$ is simply determined by equation (8).

$$\zeta_{far} = \sum_{i=1}^{U^{far}} \log_2(1 + \Gamma_{far}^i) \qquad (\frac{bit/sec}{Hz}) \quad (8)$$

Where $U^{near}$ and $U^{far}$ refer to the total number of near and far field users, respectively, which are determined by classifying the users according to the Rayleigh distance given earlier in this paper by equation (1). It is worth mentioning that the beamforming vector $w^i$ that is applied to the massive MIMO technology, in both near/far field cases is adopted according to one of the following proposed schemes.

## 4. The Proposed Beamforming Strategies

This section presents the results obtained by adopting the proposed cognitive-NOMA and NOMA inspired beamforming schemes within the considered massive MIMO-NOMA system model. Then the best approach will be adopted in the latter parts of the simulations within this paper.

### 4.1 The Proposed NOMA Inspired Beamforming Strategy

The proposed NOMA beamforming strategy is inspired by the non-orthogonal concept of NOMA. For instance, given that the $i$-th user receives the channel vector $h^i \in \mathbb{C}^{N \times 1}$. The beamforming of the near field users will be obtained according to equation (9).

$$w_{near}^i = J_{near}^i / \|J_{near}^i\|^2 \quad (9)$$

Where

$$J_{near}^i = \frac{|(h_{near}^i)^H|^2}{\sum_{m \neq i}^{U} |(h_{near}^i)^H|^2 - \sum_{j \in U^{far}} |(h_{near}^i)^H|^2 + \sigma^2} \quad (10).$$

The expression in (10) is clearly inspired by the SINR expression given in equation (4). The expression in (10) determines the beamforming vector $J_{near}^i$ of the near field users, that is characterized as interference adjusted vector. Calculating $J_{near}^i$ is performed by summation of the interference from other users as well as the noise effect, all of which is scaled by the received signal strength from the $i^{th}$ user. In such cases, each user's beam is being directed to magnify their respective signal to interference plus noise ratio (SINR).

It must be noted that the expression in equation (10) depends upon the availability of CSI at the base station, which is assumed to be known through feedback from the communicating users to the base station. By examining the expressions in (9) and (10), determining the beamforming for the near field users implies calculating the vector $J_{near}^i$ as expressed in (10), this is done by applying scalar products' summation between the $i^{th}$ near field user's channel and the channels of the other users. Given that this is applied over $U$ users, then the computation scale is $O(N \times U)$ for each single user. When calculating the beamforming vector, $w_{near}^i$, it requires normalizing the vector $J_{near}^i$ which leads to adding extra degree of complexity by $O(N)$ per each single user. Therefore, for all the near field users, the complexity would be $O(N^2 \times U)$, this is considered to be effective for modern sized systems as well as the predicted sizes of future systems.

On the other hand, the beamforming vector of the far field users is obtained through applying projection into the null space of their near field peers. This is important to nullify the cancellation of any interference between the near field and far field users while allowing excess use of the available resources through NOMA. Mathematically, the beamforming vector of the far field users could be expressed as in equation (11).

$$w_{far}^i = \epsilon(\mathbf{1} - [w_{near}^i inv((w_{near}^i)^H w_{near}^i)(w_{near}^i)^H]) \quad (11)$$

The expression in (11) generates the beamforming vector of the far field users by applying projection upon the orthogonal complement of the space of the near field users' beam. To express (11) in matrix form, it is given as in (12).

$$W_{far} = \epsilon\left(\mathbf{I} - \left[W_{near} inv\left((W_{near})^H W_{near}\right)(W_{near})^H\right]\right). \quad (12)$$

Where $\epsilon$ is used to refer to the imperfections of the SIC and channels CSI at the base station, and **inv**(…) refers to the matrix inverse operation. The matrix form of the near field users' beamforming could be expressed in a similar way to that used in equation (12), keeping in mind that in both cases the matrix is constituted of multiple vectors $W_{far} = [w_{far}^1, w_{far}^2, \dots\dots\dots w_{far}^N]$. By examining the expressions given by (11) and (12), determining the beamforming vectors of the far field users is done by performing projection onto the null space of the near field users. As given in (12), this procedure requires applying matrix inverse which possess a complexity around $O(N^3)$. In case of large-sized antenna arrays, this overhead is mitigated through applying users' clustering and also limiting the null space projection to smaller groups so as to diminish the effective complexity.

## 4.2 The Proposed Cognitive-NOMA Beamforming Strategy

The beamforming is applied in the considered massive MIMO NOMA system to focus the beam upon a given cluster of near field or far field users, independently.
In this paper, the first proposed beamforming scheme is a cognitive principled scheme that works by prioritizing the near field users as they supposedly have the highest channel powers as compared to the far field users. According to the cognitive NOMA beamforming, the far field users will be assigned the spatial resources that remain unexploited by the near field users. To obtain the beamforming vector of the near field users according to cognitive-NOMA beamforming, the expression is given in equations (13) and (14) for the near field and far field users, respectively.

$$w_{near}^i = J_{near}^i \cdot inv((J_{near}^i)^H J_{near}^i) \quad (13)$$

Where $J_{near}^i$ is given earlier by equation (10). On the other hand, the far field user's beamforming vector is obtained by (14).

$$w_{far}^i = PROJ(J_{near}^i) \quad (14)$$

The expression in equation (14) shows the cognitive-NOMA beamforming of far field users is obtained by implementing the projection of $J_{near}^i$ to obtain the orthogonally complementing the NOMA channel concept in $J_{near}^i$. From (13) and (14), it could be seen that the principles of the proposed cognitive-NOMA scheme are based on prioritizing the near field users by allocating high power beams to them. Hence, calculating $J_{near}^i$ for these users include performing inversion and matrix multiplications process which requires a complexity of around $O(N^3)$. On the other hand, from (14), it is obvious that determining the beamforming vectors for the far field users involves projection which reduces the complexity to be solely $O(N^2)$, especially when the group size is being diminished by clustering.

## 5. Connectivity Enhancement Approaches

Establishing robust connection links is vital for 6G applications, especially in densely populated environments. As this paper concerns near/far field users' wellbeing, the focus will be on this matter. There are several ways to enhance the connectivity of near/far field users. For instance, the allocated transmission power could contribute to this purpose, where proper transmission power control could reduce the effect of interference caused by near field users onto the far field users. In this regard, this paper proposes the concept of dynamic adjustment for the transmission power of near field users in accordance with the time variations of the communicating channel which maintains continuous and robust connectivity. For far field users' case where they are typically affected by sever channel conditions and path loss, this paper proposes allocating high-power levels to far users to balance and cancel the channel effect as it is possible and boost the connectivity level. One of the additional measures for future work is to adopt optimal power allocation. Another solution that could be exploited to enhance connectivity of near/far field users is to apply scheduling techniques. For near field users, this paper proposes two approaches to examine the impact that scheduling could have upon connectivity, namely, prioritized scheduling and dynamic (adaptive) scheduling. According to the prioritized scheduling concept, the near field users are favored in terms of resource allocation priority as they have high channel powers, and this scheme would lend them leverage those channel condition to reduce waste of scares resources. The concept of dynamically scheduling the users encompasses that the resources are adaptively assigned among the users according to their channel conditions and traffic demand. On the other hand, this paper proposes joint scheduling and fairness scheduling concepts for the far field users. This paper proposes joint scheduling for far field users to optimize their performance along with the near field users' performance in coordination so as to reduce the interference caused by the latter upon the former, and to balance the overall system load. This paper proposes the concept of fair scheduling to maintain a good balance the resource allocation among the near/far field users. Radio resource allocation is also a concept that could be adopted to enhance the connectivity of near/far field users. Adopting resource allocation techniques such as dynamic resource allocation and resource pooling is also effective in enhancing connectivity.

### 5.1 The Proposed Hybrid User-Clustering Approach
In this paper, it is assumed that the available radio resources are dynamically allocated within each cluster by taking into consideration the real-time variations of the channel conditions. The allocated power, in particular according to the proposed FG algorithm, is optimized and allocated per each near and far field to improve the overall energy efficiency, sum rate, and the connectivity level.
The proposed Hybrid User-clustering scheme consists of DBSCAN and spectral-clustering techniques. The first step is to cluster the users according to the spectral-clustering concept by gathering users with similar or nearby channel powers, have similar mobility profiles, and similar traffic demand levels. After that, DBSCAN clustering is performed within the primary clusters to refine the first stage outcome and to add extra flexibility in the size and shape of the resulting clusters. By taking into consideration users' clustering, the proposed beamforming approaches in this paper tend to achieve a compromise between complexity and scalability. This is possible because applying clustering minimizes the required number of calculations for matrix operations and that of the null space projections, in particular, such requirements for 6G large size implementations. Moreover, exploiting the proposed cognitive-NOMA beamforming for prioritized near filed user group minimizes the demands for real-time processing especially for less critical users and hence, diminishes the scalability in dense 6G's deployments.

### 5.2 The Proposed Fairness-Gradual Power Allocation (FG)

To achieve a compromise between the overall SE and the resource allocation fairness among the active users, this paper also proposes a fairness motivated power allocation that contributes to fair allocation of the transmission power among the users while enhancing the overall sum rate. The concept of FG encompasses gradual allocation of user's transmission power while considering maximizing the sum rate and maintaining non-zero and non-negative power allocation. Mathematically, the steps of performing FG are summarized as in Table II.

**Table II: The steps of performing the proposed FG technique**

- **Initialize** each user transmission power $P_{u,n}^s$ to be equally valued among all users, where at $s = 0$, $P_{u,n}^0 = {P^T}/{U}$.

- **Calculate** $r_u^{ff}$ for each user using $P_{u,n}^s$ at the current gradual step.

- **Update** the initial value of the transmission power, $P_{u,n}^s$, is gradually updated by $P_{u,n}^{s+1} = P_{u,n}^s + \ell \left( {r_u^{ff}}/{\sum_{u=1}^{U} r_u^{ff}} - {P_{u,n}^s}/{P^T} \right)$, where $\ell$ refers to a gradual weight parameter, the variable $s$ represents the gradual step index.

- **Check** if $\sum_{k=1}^{K} P_{u,n}^s \leq P^T$, otherwise, normalize the allocated power $P_{u,n}^s$ by applying $P_{u,n}^s = {P_{u,n}^s \cdot P^T}/{\sum_{k=1}^{K} P_{u,n}^s}$.

From Algorithm II, it is obvious that the proposed FG allocates the user's power gradually depending on the throughput achieved by each single user while maintaining the total and positive power constraints, respectively.

### 6. Simulation and Results Discussions

To guarantee that the obtained results represent a reflection of typical 6G conditions, the simulation environment, performance metrics, and parameters were chosen with, for instance, a Rayleigh flat fading channel model to emulate the effect of multipath fading, as in urban environments. The simulated system model incorporates additional environmental phenomena, such as static and moving users to accurately model user mobility, which is treated as a random variable to produce real-time results.  The considered model takes varying traffic demand levels to emulate practical scenarios with varying load performance.

In addition to the sum rate and the connectivity, energy efficiency is also used as another performance benchmark in this paper. It is calculated for the near field $\tau^{near}$ and far field $\tau^{far}$ users, according to equations (15) and (16), respectively.

$$\tau^{near} = \zeta_{near} \Big/ \sum_{i=1}^{U^{near}} p_{near}^i \qquad (15)$$

$$\tau^{far} = \zeta_{far} \Big/ \sum_{i=1}^{U^{far}} p_{far}^i \qquad (16)$$

### 6.1 Simulation Results Part 1: The Proposed Beamforming and Scheduling Schemes of Near Field Users

This section includes the proposed scheduling concepts to be performed for the near field users along with the proposed cognitive-NOMA and NOMA inspired beamforming schemes in addition to the proposed power allocation FG technique.

1. **Priority Scheduling of Near field users**

In this paper, priority scheduling is applied to maximize the overall energy efficiency and the overall sum rate of the near field users. The user's priority is determined according to its channel gain as compared to its near field peers. The channel gain of each user is considered to reflect the level of service demand of that user. Scheduling the near field users according to the priority concept is done by calculating the SINR according to equation (17).

$$\Gamma_{near}^{i-priority} = \frac{p_{near}^{total}\left|(h_{near}^i)^H w_{near}^i\right|^2 \Big/ U^{near}}{\left(\sum_{m \neq i}^{U^{near}} p_{near}^{total} |h_{near}^i w_{near}^m|^2 \Big/ U^{near}\right) + \sigma^2} \qquad (17)$$

Where $p_{near}^{total} = \sum_u^{U^{near}} p_{near}^u$, which stands for the total transmission power allocated to the near field users. After determining the SINR according to the expression in equation (17), the sum rate and the energy efficiency are calculated according to equations (5) and (15), respectively.

2. **Dynamic Scheduling of Near field users**

In this case, the near field users are dynamically adjusting their allocated transmission power in response to the instantaneous channel states to boost the sum rate and energy efficiency. The calculation of the SINR according to the dynamic scheduling scheme is performed according to equation (18).

$$\Gamma_{near}^{i-dynamic} = \frac{(p_{near}^{total} \cdot w_{near}^i)\left|(h_{near}^i)^H w_{near}^i\right|^2 \Big/ U^{near}}{\left(\sum_{m \neq i}^{U^{near}} (p_{near}^{total} \cdot w_{near}^m) |h_{near}^i w_{near}^m|^2 \Big/ U^{near}\right) + \sigma^2} \qquad (18)$$

As with the near field users' case, after determining the SINR according to the expression in equation (10), the sum rate and the energy efficiency are calculated according to equation (5) to equation (15).

The simulation results of the above concept are depicted in figures 1 and 2 in terms of the achievable sum rate and energy efficiency, respectively. These figures show the results of adopting both proposed beamforming schemes, the cognitive-NOMA and the NOMA inspired techniques. It is worth mentioning that the power is allocated according to the proposed FG power allocation scheme.

Figure 1 shows that dynamic scheduling offers better sum rate performance trends than the priority concepted scheduling. Despite both techniques being dependent on the channel conditions of the near field users, both metrics' results show that the dynamic approach shows better trends as it enables the users to cope up with the environment variation better than its priority concepted counterpart. In terms of the proposed beamforming schemes, the cognitive-NOMA beamforming shows a higher sum rate for the near filed users than the case of NOMA inspired beamforming scheme. This is also the case for the energy efficiency results that are illustrated in figure 2. The dynamic scheduling also shows better performance than the priority scheduling case and so is the cognitive-NOMA beamforming.

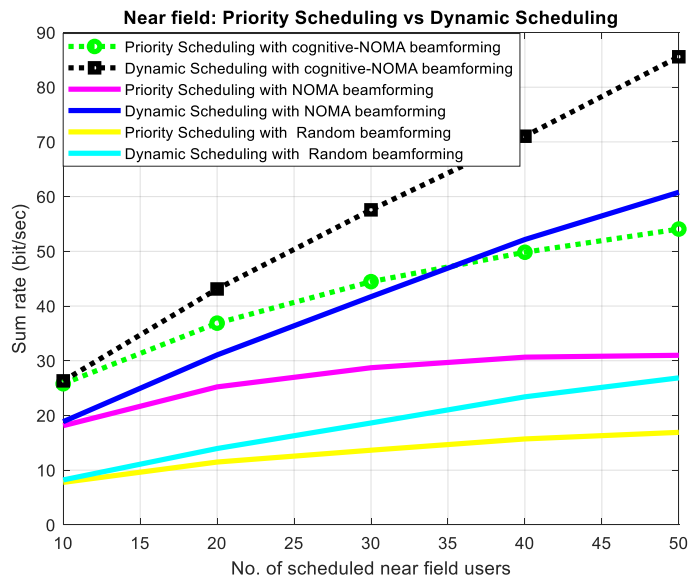

**Fig. 1:** The achieved sum rate of near field users with the proposed FG power allocation, for the priority and dynamic scheduling cases using both proposed beamforming schemes.

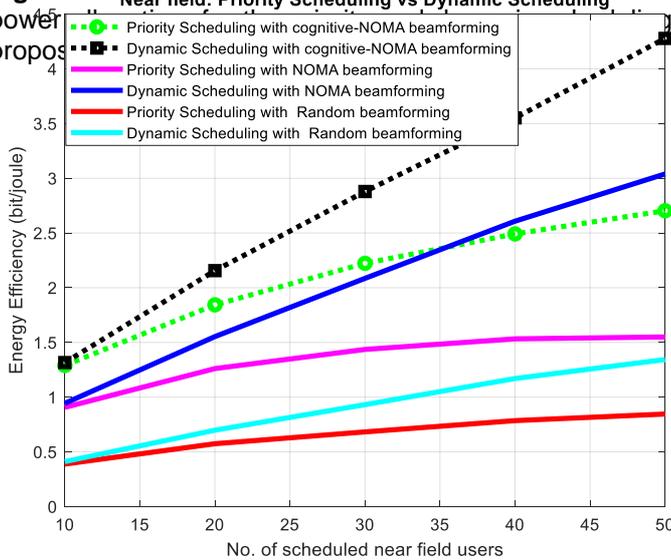

**Fig. 2:** The achieved energy efficiency of near field users with the proposed FG power allocation, for the priority and dynamic scheduling cases using both proposed beamforming schemes.

## 6.2 Simulation Results Part 2: The Proposed Beamforming and Scheduling Schemes of Far Field Users

1. **Fair Scheduling of Far Field Users**

This concept ensures that far field users get fair radio resource allocation as compared to the near field users. This is important to avoid unbalanced resource allocation between near field and far field users and also to avoid degradation of their performance. This concept tends to allocate the transmission power nearly in an equal manner among the far field users to maintain semi-fair quality of service among them. Hence, according to this concept the expression of the received SINR is given as in equation (19).

$$\Gamma_{far}^{i-fair} = \frac{p_{far}^{total}\left|\left(h_{far}^{i}\right)^{H}w_{far}^{i}\right|^{2}/U^{far}}{\left(\sum_{m\neq i}^{U^{far}} p_{far}^{total}\left|h_{far}^{i}w_{far}^{m}\right|^{2}/U^{far}\right) + \sigma^{2}} \quad (19)$$

Where $p_{far}^{total} = \sum_{u}^{U^{far}} p_{far}^{u}$ denotes the total transmission power allocated to the far field users. It must be noted that $p_{far}^{total} + p_{near}^{total} = p^{total}$. The expression in equation (19) is then subbed in equation (8) and equation (16) to determine the sum rate and the energy efficiency of the far field users, respectively.

2. **Joint Scheduling of Far Field Users**

Unlike the fair field scheduling concept, the joint scheduling has the idea of allocating the resources and the transmission power in a biased manner among the far field users in accordance with their channel gains. In this manner, the scheduling process would be more adaptive to the peripheral environment that would result in SINR being received at each far user matching the expression of equation (20).

$$\Gamma_{far}^{i-joint} = \frac{\left(p_{far}^{total}.w_{far}^{i}\right)\left|\left(h_{far}^{i}\right)^{H}w_{far}^{i}\right|^{2}/U^{far}}{\left(\sum_{m\neq i}^{U^{far}} \left(p_{far}^{total}.w_{far}^{m}\right)\left|h_{far}^{i}w_{far}^{m}\right|^{2}/U^{far}\right) + \sigma^{2}} \quad (20)$$

The achievable sum rate and energy efficiency of the far users is calculated by replacing the expression of equation (20) into equation (8) and equation (16), respectively. Figure 3 and figure 4 illustrate the sum rate and energy efficiency of the joint and fairness-based scheduling of far filed users in the downlink of massive MIMO NOMA system. The results were obtained using the proposed FG power allocation scheme along with both of the beamforming and scheduling techniques. These figures show that the system considered works better with joint scheduling techniques as compared to the case where it adopts a fairness-based approach. From these figures, it could be seen that the fairness approach tends to keep stable behavior along with the changing numbers of users, which in this case, represents the load and the traffic demand rate. On the other hand, joint scheduling benefits from the increasing numbers of users to make profit in terms of users' diversity gain which is also supported by the availability of multiple antennas at the base station. Unlike the near field users' case, the proposed NOMA inspired beamforming shows comparable results to that of the proposed cognitive-NOMA scheme, with less complexity.

This is evidence that the proposed NOMA beamforming scheme could be adopted for far field users and the proposed cognitive-NOMA scheme to be adopted for near field users.

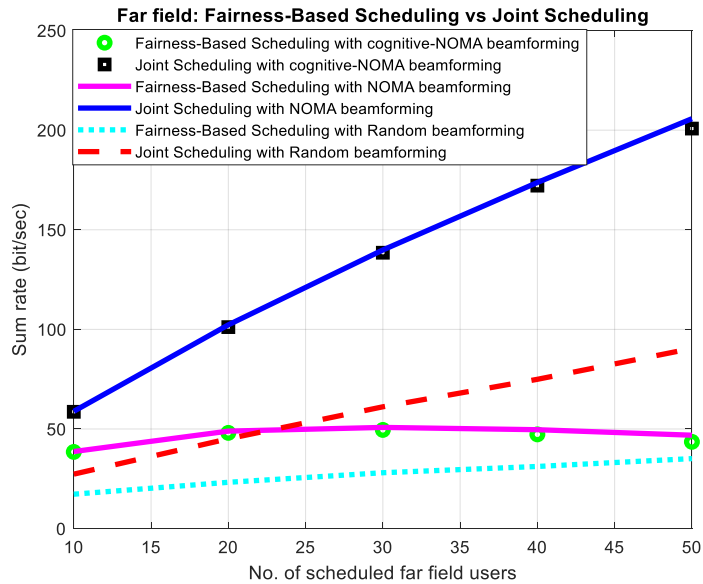

**Fig. 3:** The achieved sum rate of far field users with the proposed FG power allocation, for the fairness based and joint scheduling cases using both proposed beamforming schemes.

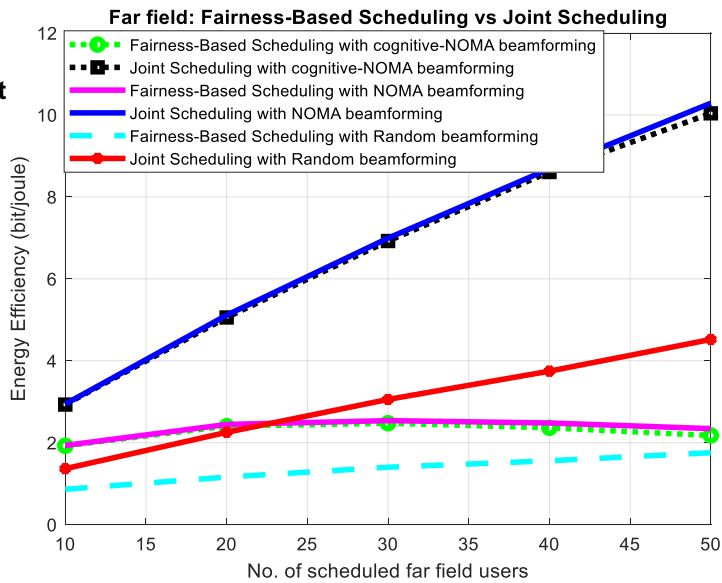

**Fig. 4:** The achieved energy efficiency of far field users with the proposed FG power allocation, for the fairness based and joint scheduling cases using both proposed beamforming schemes.

### 6.3 Simulation Results Part 3: The Proposed cognitive-NOMA beamforming and the Impact of Clustering Upon Scheduling in Near/Far Field Users

The third criterion to examine in this paper is the level of improvement gained upon the connectivity, sum rate, and energy efficiency in case of no clustering applied against the clustering-based case. The results obtained within the previous part proved that the proposed cognitive-NOMA beamforming showed better results than the NOMA inspired scheme, hence, it will be adopted here while obtaining the results of this part of the paper along with the proposed FG power allocation algorithm.

Figures 5 and 6 display the connectivity of near field and far field users, respectively, using the proposed beamforming and power allocation schemes. Both figure 5 and figure 6 show that applying clustering helps significantly in enhancing the level of connectivity provided as compared to the case of no clustering. This is mainly because clustering simplifies the scheduling process and boosts the connectivity as well as the level of the provided coverage to the communicating users. In addition to clustering, it is crucial to highlight the spatial gain acquired by exploiting massive MIMO at the base station where a total of 128 antennas were used to create spatial gain and help serve the active users.

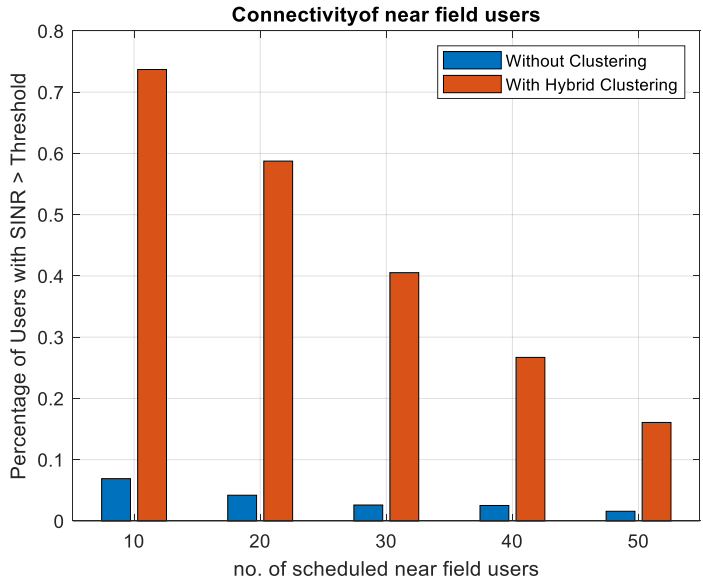

**Fig. 5:** The impact of clustering on the connectivity performance of the near users with total transmission power of 20W. With the application of the proposed cognitive-NOMA beamforming and the proposed FG algorithm.

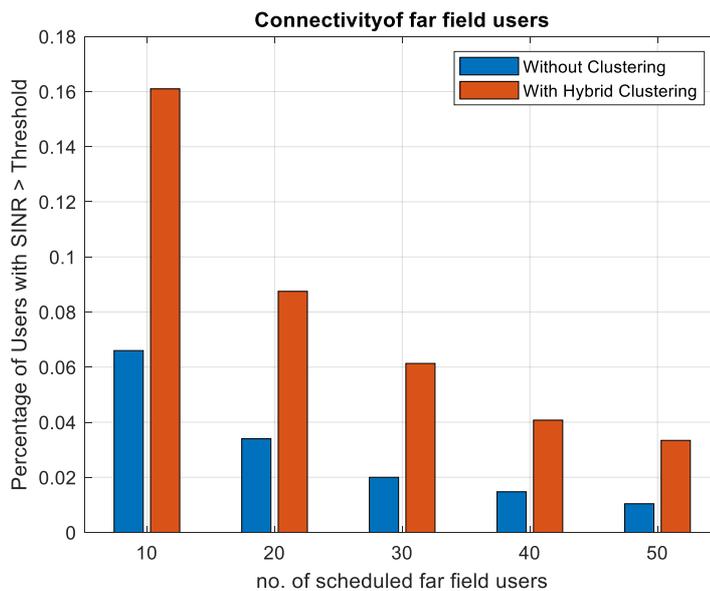

**Fig. 6:** The impact of user-clustering on the connectivity performance of the far users with total transmission power of 20W. With the application of the proposed cognitive-NOMA beamforming and the proposed FG algorithm.

Figures 7 and 8 respectively display the sum rate of the near field and the far field users with the proposed FG allocation scheme and cognitive-NOMA beamforming. The obtained sum rates confirm that clustering offers better performance and plays an important role in scheduling and overall performance. It is important to highlight the gain acquired from the spatial dimension of the massive MIMO and the eff

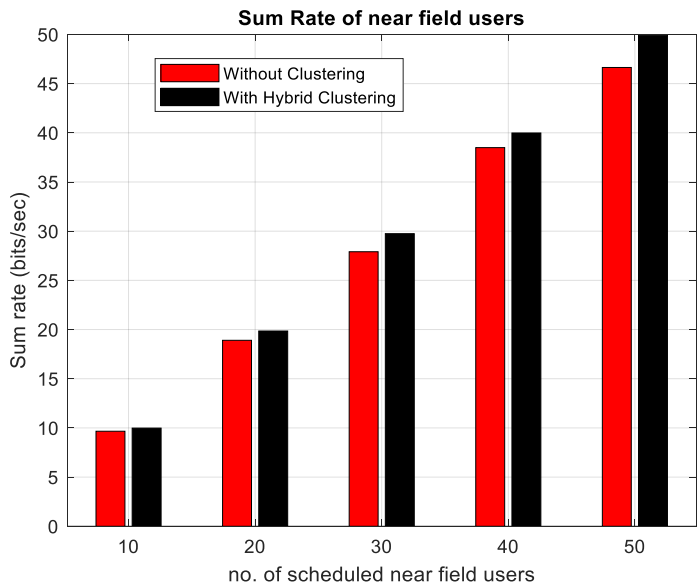

**Fig. 7:** The impact of clustering on the performance of the near users in terms of the achieved sum rate.

icient resource utilization gained from applying NOMA. All those factors contributed to the significant performance improvement for the case of clustering as compared to the no clustering case. These figures, again, are confirmation that clustering enhances the far field users' experience which is necessary for the envisi

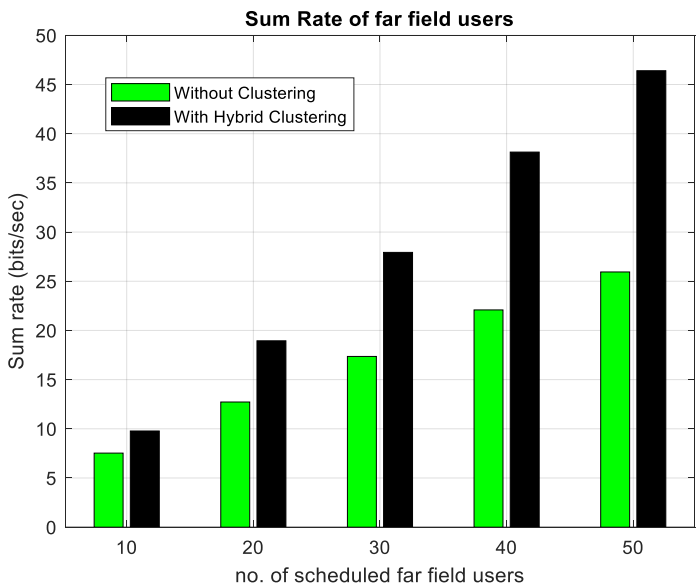

**Fig. 8:** The impact of clustering on the performance of the far users in terms of the achieved sum rate.

oned requirements of future wireless systems.

Figures 9 and figure 10 explore the effect of clustering upon the near/far users' performance at the massive MIMO-NOMA system in terms of energy efficiency. using the proposed FG power assignment and the proposed cognitive-NOMA beamforming technique. There is an obvious advantage for the case that adopts user-clustering as compared to the no clustering case. It means that clustering the users helps in allocating the power at higher accuracy because it reflects the real need of the users to the transmission power taking into consideration important propagation factors such as the channel power, mobility and traffic demand of each user. This is also a positive benefit in favor of clustering applications as it facilitates

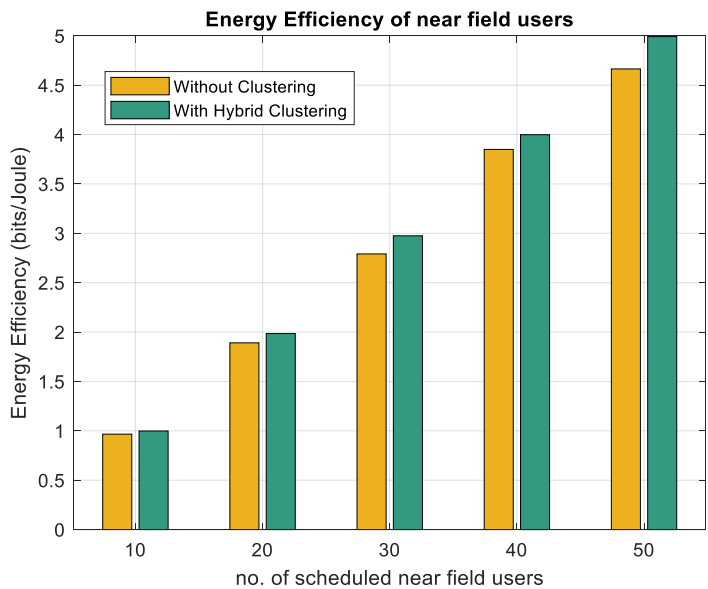

**Fig. 9:** The impact of user-clustering on the performance of the near users in terms of the achieved energy efficiency using the proposed FG power assignment and the proposed cognitive-NOMA beamforming technique.

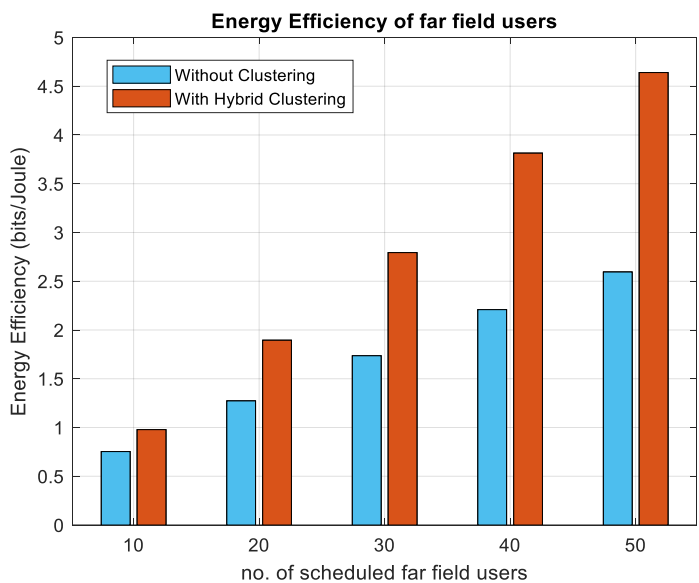

**Fig. 10:** The impact of user-clustering on the performance of the far users in terms of the achieved energy efficiency.

applying FG and user scheduling which return positively upon the overall system performance.

## 7. Conclusions and future work

This paper addressed the impact of users clustering on the performance of near/far field users and examined for a downlink massive MIMO-NOMA system taking into consideration optimizing the power allocated among the users using the proposed FG power allocation algorithm. In addition, this paper also proposes two beamforming schemes that are inspired by the NOMA concept. The obtained results, in general, show that clustering the users would not just facilitate dealing with their predicted large number, but it also proves its effectiveness in dealing with densely active users' scenarios that are envisioned in the future networks and the results also prove that clustering simplifies service fulfillment. Regarding the simulation results related to the proposed beamforming strategies, a comparison against random beamforming is included to examine the performance of the proposed schemes against a competing contender from literature. The obtained results show that the proposed cognitive-NOMA beamforming showed better performance than its proposed contender that is based solely on NOMA channels. However, the latter requires less complex mathematical operations than the latter. The near field users result also showed better results using NOMA inspired beamforming as compared to the cognitive-NOMA beamforming. This might draw a conclusion to adopt cognitive-NOMA beamforming for far field users while adopting NOMA beamforming for near field user's scenarios. For the simulation part related to examine the clustering impact upon the considered system performance, the proposed cognitive-NOMA beamforming scheme is adopted as it proved its privilege over the proposed NOMA beamforming approach. The results obtained in this part show that a near/far massive MIMO-NOMA system would have higher sum rate, energy efficiency, and better connectivity than the case where no user clustering is applied. This is because without clustering, the available resources wouldn't benefit from as with the case of clustering, especially since clustering the users who have similar channel conditions, and similar mobility and traffic demand profiles. All in all, the proposed approaches are considerably feasible for real-time applications as they maintain hierarchical computations structure through clustering. This is proof that it would provide support for the predicted scalability in large 6G networks.

**Declaration Statements:**
**Funding:** This research did not receive any specific funding
**Conflict of Interest:** The authors declare no conflict of interest
**Acknowledgements:** I would like to express my very great appreciation to the co-authors of this manuscript for their valuable and constructive suggestions during the planning and development of this research work.
**Informed consent:** Not Applicable
**Ethical approval:** Not Applicable
**Author Contribution:** All authors have made substantial contributions to conception and design, revising the manuscript, and final approval of the version to be published. Also, all authors agreed to be accountable for all aspects of the work in ensuring that questions related to the accuracy or integrity of any part of the work are appropriately investigated and resolved.
**Data Availability Statement:**
No new data was generated or analyzed in support of this research.

Optimized Connectivity Through Sched.                                                                                                                      20